\documentclass[prl,twocolumn,showpacs,superscriptaddress]{revtex4}
\usepackage{amsmath,amssymb,mathrsfs}
\usepackage[dvips]{graphicx}
\usepackage{hyperref}
\usepackage{paralist}

\begin{document}

\title{Integrability vs exact solvability in the quantum Rabi and Dicke models}

\author{Murray T. Batchelor}
\email[e-mail:]{batchelor@cqu.edu.cn}
\affiliation{Centre for Modern Physics, Chongqing University, Chongqing 400044, China}
\affiliation{Department of Theoretical Physics,
Research School of Physics and Engineering,
Australian National University, Canberra ACT 0200, Australia}
\affiliation{Mathematical Sciences Institute, Australian
National University, Canberra ACT 0200, Australia}

\author{Huan-Qiang Zhou}
\email[e-mail:]{hqzhou@cqu.edu.cn}
\affiliation{Centre for Modern Physics, Chongqing University, Chongqing 400044, China}

\begin{abstract}
The Rabi model describes the simplest interaction between light and matter via a two-level quantum system
interacting with a bosonic field. 
We demonstrate that the fully quantised version of the Rabi model is integrable in the Yang-Baxter sense at two parameter values.
The model is argued to be not Yang-Baxter integrable in general. 
This is in contrast to the claim that the quantum Rabi model is integrable based on a 
phenomenological criterion of quantum integrability not presupposing the existence of a set of commuting operators. 
Similar Yang-Baxter integrable points are identified for the generalised Rabi model and the fully quantised Dicke model.
The integrable points have particular implications for the level statistics of the Dicke model.
\end{abstract}

\maketitle

Integrability is arguably the most powerful concept in the mathematical description of physical systems. 
A classical system is defined to be integrable when the number of degrees of freedom 
is smaller than the number of independent constants of the motion \cite{classical}. 
However, for quantum systems, the definition of integrability is not so clear \cite{JS,Larson}. 
Among the various definitions, 
the concept of Yang-Baxter integrability \cite{McGuire,Yang,Baxter} is particularly powerful and 
seems most appropriate for (1+1)-dimensional quantum systems.
Solutions of the Yang-Baxter relation along with associated monodromy matrices allow 
the construction of integrable models and their conserved charges \cite{Korepin}, 
from which physical information can be derived exactly. 
Indeed Yang-Baxter integrable (YBI) models are synonymous with the term exactly solved models \cite{Baxterbook}.
Nevertheless not all exactly solved models are necessarily YBI. 
Moreover, YBI is arguably a not necessary but sufficient condition to guarantee integrability.

A general criterion of quantum integrability, inspired by the classically integrable hydrogen atom and 
not requiring the existence of a set of 
commuting operators, has been proposed  
in the context of the quantum Rabi model \cite{Braak1}.
The Rabi model \cite{Rabi} describes a two-level quantum system
interacting with a bosonic field. 
It models the simplest interaction between light and matter and is thus 
a fundamental model in quantum physics. 
Applications include the interaction between light and trapped ions or quantum dots \cite{qdots} 
and between microwaves and superconducting qubits \cite{sbits}. 
The Rabi model is applicable to both cavity \cite{cavity} and circuit \cite{circuit} QED.

Despite it's simplicity, the fully quantised version of the Rabi model was solved only recently 
and claimed to be integrable \cite{Braak1}.
Briefly stated, Braak's criterion of quantum integrability, involving $f_1$ discrete and $f_2$ continuous 
degrees of freedom, 
is that 
integrability is equivalent to the existence of $f=f_1+f_2$ ``quantum numbers'' to classify eigenstates uniquely.
For the Rabi model $f_1=f_2=1$, giving $f=2$ which is the same dimension as the global label (parity) used to 
uniquely label the eigenstates. 
The criterion demands that the number of values which the label tied to the discrete degree of freedom can take 
equals the dimension of the corresponding Hilbert space.
This condition is satisfied for the Rabi model.

A natural question arises: if the quantum Rabi model is integrable -- {is it YBI?}
Here we show that the quantum Rabi model is YBI at two distinct   
parameter values.    
Significantly, the Rabi model does {not} appear to be YBI in general. 
This raises a question with regard to the utility of Braak's criterion of quantum integrability, 
which we further discuss here.
We also identify corresponding YBI points in 
the Dicke model \cite{Dicke}, which is the extension of the Rabi model to $N$ qubits, 
a model also of fundamental interest.
For $N=2$ the Dicke model constitutes a simple model of the universal quantum gate \cite{qgate}.
It may also be possible to realise the 
$N=3$ Dicke model within circuit QED \cite{casanova}.
Most recently an analog-digital quantum simulation for all parameter regimes of the quantum Rabi and Dicke models 
 has been proposed using circuit QED  \cite{sim}. 
The Dicke model is also of interest for large $N$ where it exhibits a phase transition to a
super-radiant state for strong coupling \cite{largeN,EB2003,BGBE}.

\textit{The quantum Rabi model.-} 
The hamiltonian of the fully quantised version of the Rabi model (with $\hbar=1$) is 
\begin{equation}
H = \Delta \, s^z + \omega \, a^{\dagger} a + g \, s^x (a + a^{\dagger}), 
\label{rabi}
\end{equation}
where $s^x$ and $s^z$ are spin-$\frac12$ matrices for the two-level system with level splitting $\Delta$.
$a^\dagger$ ($a$) denote creation (destruction) operators for a single bosonic mode with 
$[a, a^\dagger] = 1$ and frequency $\omega$.
$g$ is the coupling between the two systems.
The quantum Rabi model has $Z_2$ symmetry (parity).

Using the representation of the bosonic operators in the Bargmann space of analytic functions, 
the regular eigenvalues of the quantum Rabi model were shown to be given in terms of the 
zeros of a function $G_\pm(x)$ \cite{Braak1,Braak2}.
Simple poles of $G_\pm(x)$ at $x=0, \omega, 2\omega, \ldots$ 
correspond to the eigenvalues of the uncoupled bosonic modes.
We will call models with solutions of this type Braak solvable.
The conditions proposed by Braak are a type of sufficiency condition for determining the regular solutions.
They also include the exceptional eigenvalues, which are the well known Juddian isolated exact solutions \cite{Judd}. 
Symmetric, anti-symmetric and asymmetric solutions for the eigenstates are 
given in terms of confluent Heun functions \cite{Huen1a,Huen1b}, which involve an infinite number of terms. 
The isolated exact solutions appear naturally as truncations of the confluent Heun functions.

The rotating wave approximation was used to treat the fully quantised version of the Rabi model (\ref{rabi}) in the form
\begin{equation}
H_{JC} =  \Delta \, s^z + \omega \, a^{\dagger} a + g \, (s^+ a + s^- a^{\dagger}), 
\label{JC} 
\end{equation}
with   $s^\pm = s^x \pm {\mathrm i} s^y$.
This is the Jaynes-Cummings (JC) model \cite{JC}. 
The conditions of near resonance $\Delta \approx \omega$ and weak coupling $g \ll \omega$ for the 
rotating wave approximation apply in many experimental settings. 
The JC model is YBI \cite{review}. 
The excitation number operator $M=a^\dagger a + s^z$ and the Casimir operator 
$s^2 = s^+ s^- + s^z(s^z-1)$ commute with hamiltonian (\ref{JC}), i.e., $[H_{JC},M] = [H_{JC},s^2]=0$.

\textit{Dicke Model.-} We consider the Rabi model in the context of the more general Dicke model \cite{Dicke}, 
for which the radiation mode couples to $N$ two-level qubits.
We write the Hamiltonian in the form ($\hbar=1$)
\begin{equation}
H_D = 2 \Delta \, S^z + \omega \, a^{\dagger} a + g \, (S^+ + S^-) (a + a^{\dagger}), 
\label{dicke2}
\end{equation}
where now 
\begin{equation}
S^z  = \sum_{j=1}^N s_j^z, \quad S^x  = \sum_{j=1}^N s_j^x, \quad S^{\pm} = \sum_{j=1}^N s_j^\pm .
\label{N}
\end{equation}
Apart from a harmless redefinition of the system parameters, the 
Rabi hamiltonian (\ref{rabi}) follows for  $N=1$.
The quantum Rabi model with two qubits \cite{Peng,Chen2} and the $N=3$ Dicke model \cite{Braak4} have 
also been shown to be Braak solvable.
According to Braak's integrability criterion, the Dicke model is non-integrable for all $N \ge 2$ \cite{Braak4}.

Applying the rotating wave approximation to the Dicke model leads to the Tavis-Cummings model \cite{TC}, with  hamiltonian
\begin{equation}
H_{TC} =  \Delta \, S^z + \omega \, a^{\dagger} a + g \, (S^+ a + S^- a^{\dagger}) .
\label{TC} 
\end{equation}
For this model the operators $M=a^\dagger a + S^z$ and  
$S^2 = S^+ S^- + S^z(S^z-1)$ commute with $H_{TC}$.
The Tavis-Cummings model reduces to the JC model for $N=1$. 
It is YBI for general $N$ and can be solved by the algebraic Bethe Ansatz \cite{review}.

We find that the Dicke model (\ref{dicke2}) is YBI for the two cases (i) $\Delta=0$ and (ii) $\omega=0$. 
In both cases, there is an extra conserved quantity $C$, i.e., $[H_R, C]=0$.
For $\Delta=0$, $C = S^{+} + S^-$, while for $\omega=0$, $C= a^\dagger + a$.
To establish Yang-Baxter integrability, the key idea we introduce is 
an operator-valued twist, 
which in this setting yields a ``trivial" twist solution to the Yang-Baxter relation.
These solutions establish the YBI of the model.

First consider the case $\Delta=0$. 
We construct the transfer matrix operator $\tau(u)={\mathrm{tr}} \, T(u)$, where the monodromy matrix 
$T(u) = W^S L^a(u)$ is a combination of the spin operator-valued ``twist"
\begin{equation}
W^S =\begin{bmatrix} 1 & S^{+}+S^{-} \\ S^{+}+S^{-}  & -1
\end{bmatrix}, \label{Ws}
\end{equation}
and the bosonic $L$-operator \cite{zhou}
\begin{equation}
L^a(u) = 
\begin{bmatrix} 1+\eta u+\eta^2 N & \eta a \\ \eta a^{\dagger} & 1
\end{bmatrix}, \label{mon1}
\end{equation}
where $\eta$ is a free parameter and $N = a^\dagger a$. 
The elements of $T(u)$ can then be shown to satisfy the intertwining relation \cite{review,zhou}
\begin{equation}
R_{12}(u-v)T_1(u)T_2(v)=T_2(v)  T_1(u) R_{12}(u-v),
\label{intertwine}
\end{equation}
with the (standard) $R$-matrix
\begin{equation}
R_{12}(u)=\begin{bmatrix} u+\eta & 0 & 0 & 0\\ 0& u & \eta & 0 \\  0 & \eta &u & 0 \\ 0 & 0 & 0&u+\eta
\end{bmatrix} , 
\label{R}
\end{equation}
satisfying the Yang-Baxter relation 
\begin{equation}
R_{12}(u-v)R_{13}(u)R_{23}(v)=R_{23} (v)  R_{13}(u) R_{12}(u-v), 
\label{YBE}
\end{equation}
which is the so-called masterkey to integrability \cite{perk}.
It follows that  
\begin{eqnarray}
\tau(u)&=&\eta[u+\eta N+     (S^{+} + S^{-}) (a^{\dagger}+a)] \nonumber\\
&=&\eta[u+g^{-1}H_D],
\end{eqnarray}
where we have identified $\eta={\omega}/{g}$.

For the case $\omega=0$ the monodromy matrix is of the form $T(u) = W^a L^S(u)$ 
with the bosonic operator-valued twist 
\begin{equation}
W^a = \begin{bmatrix} 1+\lambda & a+a^+\\ a+a^+&1-\lambda
\end{bmatrix} , \label{Wa} 
\end{equation}
where $\lambda={\Delta}/{g}$ and the spin $L$-operator is \cite{review,zhou}
\begin{equation}
L^S(u) = \begin{bmatrix} u+\eta S^z & \eta S^{-} \\ \eta S^+ &u-\eta S^{z}
\end{bmatrix} , \label{mon2}
\end{equation}
where $\eta$ is a free parameter.
$T(u)$ defined in this way can also be shown to satisfy the intertwining relation (\ref{intertwine}) with the $R$-matrix (\ref{R}).
In this case an alternative factorised form of the monodromy matrix is 
\begin{equation}
T(u)= W^a \prod^N_{j=1} L_j^s(u), \label{higher}
\end{equation}
where the spin operator $L_j^s(u)$ is defined in an obvious way from the single spin term for each site $j$.
In this case the operator $\tau(u)={\mathrm{tr}} \, T(u)$ is a polynomial of degree $N$, with 
\begin{equation}
\tau(u) = 2 u^N + \eta g^{-1} u^{N-1}  H_D + \ldots \,.
\label{poly}
\end{equation}

For the two YBI cases $\Delta=0$ and $\omega=0$ 
we have thus obtained the commuting operators $[\tau(u),\tau(v)]=0$. 
The bosonic and spin operators $L^a(u)$ and $L^S(u)$ appearing in the monodromy matrices 
are standard forms, the new ingredients in each case are 
the corresponding operator-valued twists $W^S$ and $W^a$.
The parameter value $\omega=0$ of the Dicke model (\ref{dicke2}) has been identified  as YBI using another approach.
In particular, a Bethe Ansatz solution has been obtained from the elliptic Gaudin model 
through a limiting procedure \cite{Kundu}.

It is well known that the Rabi and Dicke models can be solved by elementary means 
at the value $\Delta=0$, the degenerate atomic limit.
At this point the model can be solved, e.g., by using a coherent-state representation \cite{EB} or a 
polaron transformation \cite{ABEB}. 
The value $\omega=0$ is also trivial since $a + a^{\dagger}$ is proportional to a position operator. 
Here we have identified these values \cite{footg} as YBI points.

We have been unable to find any solution which interpolates between these two YBI cases. 
Indeed, we believe there is no such solution, an argument for which can be made as follows.
For the Rabi model, recall that the essential extra conserved quantity is $C = s^x$ for 
$\Delta=0$ and $C= a^\dagger + a$ for $\omega=0$.
Consider a parametrisation interpolating between the integrable points such that $\Delta = r \sin \theta$ and 
$\omega = r \cos \theta$ for some $r$. 
The desired more general quantity $C$ must satisfy $[H_R, C]=0$ and reduce to (i) $C=s^x$ for $\theta=0$ and 
(ii) $C= a^\dagger + a$ for $\theta = \frac{\pi}{2}$.
First consider the limit $\theta \to 0$, it suffices to write $H_R = H_0 + r \sin \theta \, s^z$ where $H_0$ is defined 
in an obvious way and  
$C = s^x + f(\theta) A$ for some arbitrary function $f(\theta)$ and operator $A$, with $f(0)=0$.
We know that  $[H_0, s^x]=0$ but the only operator $A$ that can ensure $[H_R, C]=0$ is $A=0$. 
Similarly consider $H_R = \widetilde{H_0} + r \cos \theta \, a^\dagger a$ with $C = a^\dagger + a + g(\theta) B$ 
for some arbitrary function $g(\theta)$ and operator $B$, with $g(\frac{\pi}{2})=0$.
Again we are led to conclude that the operator $B=0$. 
Any other situation or combination of operators would be highly unusual. 

It should also be noted that attempts have been made to include extra terms beyond the rotating wave approximation  
in the JC model in order to preserve Yang-Baxter integrability \cite{Frahm}.
These attempts, although yielding interesting results, were also unable to realise the quantum Rabi model.

\textit{Generalised Rabi model.-} Our approach using operator-valued twists may be 
applied to other models. Consider the generalised Rabi model
\begin{equation}
H_\epsilon = 2 \Delta \, s^z + \omega \, a^{\dagger} a + g \, (s^+ + s^-)  (a + a^{\dagger})   + \epsilon \, s^x ,
\label{gen}
\end{equation}
where the additional term $\epsilon \, s^x$ allows tunnelling between the two atomic states. 
It breaks the parity symmetry. 
This model is relevant to the description of hybrid mechanical systems \cite{hybrid,Huen2} 
and is referred to as the driven Rabi model \cite{Larson}.
It is also Braak solvable \cite{Braak1}, with eigenstates given in terms of Heun functions \cite{Huen2}.

The generalised Rabi model (\ref{gen}) is considered to be non-integrable \cite{Braak1}. 
The argument is that $H_\epsilon$ has no discrete symmetry and there is only one quantum number (energy) corresponding to the sole 
conserved quantity. Since the number of degrees of freedom exceeds one the model does not satisfy Braak's criterion of quantum integrability. 
It is thus claimed to be the first example of a non-integrable but exactly solvable system \cite{Braak1}.
However, one can also construct YBI points for this model at the parameter values $\Delta=0$ and $\omega=0$.
To do this we need only extend the operator-valued twist matrices.

For the case $\Delta=0$, we define the monodromy matrix $T(u) = W^s L^a(u)$, 
where 
\begin{equation}
W^s =\begin{bmatrix} 1 & 2 s^x \\ 2 s^x  & -1 + b \, s^x
\end{bmatrix}, 
\end{equation}
with $L^a(u)$ as defined in (\ref{mon1}). 
Here we have the commuting operator $\tau(u)=\eta[u+g^{-1} H_\epsilon]$, with $\eta=\omega/g$ and $b=\epsilon \eta/g$.
Similarly for $\omega=0$, we take $T(u) = W^a L^s(u)$, with  
\begin{equation}
W^a = \begin{bmatrix} 1+\lambda & a+a^+ + c\\ a+a^++c&1-\lambda
\end{bmatrix} , 
\end{equation}
and $L^s(u)$ as defined in (\ref{mon2}).
Here $\tau(u) = 2 u + \eta g^{-1} H_\epsilon$, with $\lambda = \Delta/g$ and $2c=\epsilon/g$.

\textit{Discussion.-} To conclude, we have found two distinct parameter values, $\Delta=0$ and $\omega=0$, 
both at which the quantum Rabi model (\ref{rabi}) is YBI.
The associated monodromy matrices and the $R$-matrix (\ref{R}) ensure YBI. 
The model does not appear to be YBI in general. 
We have also demonstrated that at each of the parameter values $\Delta=0$ and $\omega=0$ 
the Dicke (\ref{dicke2}) and generalised Rabi (\ref{gen}) models are YBI.
This result may also be extended to the Dicke version \cite{EB} of the generalised Rabi model (\ref{gen}).
The underlying $R$-matrix (\ref{R}) is seen to be a common feature of integrability for these and related \cite{zhou,review} models.

The existence of YBI points has various implications for these models. 
The result (\ref{poly}) following from the monodromy matrix (\ref{higher}) 
for the value $\omega=0$ leads to the 
construction of a series of higher order conserved quantities with increasing $N$. 
The presence of these conserved quantities implies level crossing, 
therefore the energy level statistics at the Dicke model integrable point 
should follow Poissonian level statistics as anticipated by the Berry-Tabor criterion \cite{bt}.
Away from an integrable point the energy level statistics should follow a Wigner-Dyson distribution. 
The quantum Rabi model appears to obey neither Poisson or Wigner-Dyson level statistics \cite{Larson,Kus}.
The situation for the Dicke model integrable point $\omega=0$ appears analogous to the Heisenberg XXZ spin-$\frac12$ chain, 
for which the expected Poissonian level statistics is not readily apparent for small chains. 
We stress that for the Heisenberg XXZ chain the existence of a classical limit is not necessary to ensure 
Poissonian level statistics. 
We thus expect that Poisson level statistics should become more apparent at or in the vicinity of the Dicke model 
integrable point with increasing $N$.
Level statistics have also been investigated extensively for the Dicke model \cite{EB2003,EB2003b,AH}.
Interestingly, for large $N$ the observed level statistics appears to be Poissonian below the super-radiant phase transition 
and Wigner-Dyson above the transition.
The identification of YBI points in these models should also be of relevance to the connection between 
integrability and thermalisation \cite{Larson}.

Our results bring into question the usefulness and validity of Braak's 
phenomenological criterion of quantum integrability \cite{Braak1}. 
The Rabi model only appears to be YBI at two points. 
If the Rabi model is not YBI in general then clearly Braak's integrability criterion does not fit with Yang-Baxter integrability. 
We have also shown that the Dicke model and the generalised Rabi model have YBI points, 
yet they are non-integrable models according to Braak's integrability criterion. 
The interplay between the notions of exact solvability, Yang-Baxter integrability and Braak solvability  
is summarised in Figure 1 in the context of the Rabi model.  
Also indicated are the exact isolated Judd Points. 
Contrary to what might be expected, the Judd points are not YBI points. 
This is because Yang-Baxter integrability deals directly with the hamiltonian of the entire system, not with the two-dimensional 
subspace associated with the Judd points.  
For given parameter values the system parameters satisfy constraint conditions at the Judd points.
These constraints are not shared by the 
regular part of the eigenspectrum, which renders the corresponding hamiltonian non-integrable.

As already noted, the Rabi and generalised Rabi models have 
been solved analytically \cite{Braak1,Chen1,Braak2,Huen1a,Huen1b,Huen2}.  
For the Rabi model, the part of the eigenspectrum corresponding to the Juddian isolated exact solutions 
can also be derived algebraically \cite{Koc} using ideas \cite{BD} based on the notion of quasi-exact solvability \cite{QES}.
The Rabi model has been called a quasi-exactly solved model \cite{quasi,zhang}.
Yet for all intents and purposes, the Rabi model {\em is} an exactly solved model, 
albeit not of the YBI kind. 
Although the term exactly solved model is often synonymous with Yang-Baxter integrability, 
this is not always the case, as indicated in Figure 1.
For YBI models, the complete eigenspectrum can be described algebraically 
in terms of finite polynomials, which is a feature, 
e.g., of finite-sized systems solved in terms of the Bethe Ansatz and related $T-Q$ relations \cite{footba}.
The solutions of the Jaynes-Cummings and the more general Tavis-Cummings models are of this form \cite{review}.
That the analytic solution of the Rabi model is not in general of this particular form lends further weight to the Rabi model 
not being YBI in general.

%%%%%%%%%%%%%%%%%%%%%%%%%%%%%%%%%%%%%%%
\begin{figure}[h]
\vspace{4mm}
\includegraphics[width=0.6\linewidth]{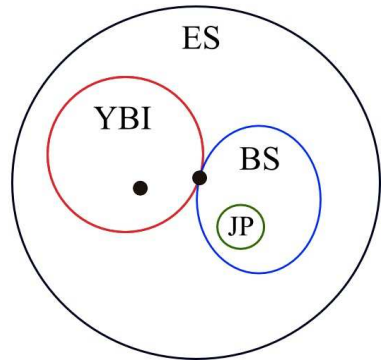}
\caption{Depiction of the relationship between  
Yang-Baxter integrable (YBI) and Braak solvable (BS) models in the context of the Rabi model. 
Each method of solvability fits within the general framework of exactly solved (ES) models. 
Also indicated are the Judd points (JP) which can be obtained from Braak's solution or via a variety 
of algebraic methods. 
According to our picture, the Rabi model, although Braak solvable, 
does not in general fit within the concept of Yang-Baxter integrability. 
The YBI and BS circles have the degenerate atomic limit $\Delta=0$ 
in common, indicated by a point in the figure. Also indicated is the other YBI point at $\omega=0$.
} \label{fig:contour}
\end{figure}
%%%%%%%%%%%%%%%%%%%%%%%%%%%%%%%%%%%%%%%

{\em Acknowledgments.} We thank Daniel Braak, Carl Bender, Michael Berry, Fabian Essler and an anonymous referee for helpful remarks. 
 M.T.B.  acknowledges support from the 1000 Talents Program of China. 
 This work is supported in part by the National Natural Science
Foundation of China (Grant No. 11174375) and by the Australian Research Council.


\begin{thebibliography}{99}

\bibitem{classical} See, e.g., H. Goldstein, {\em Classical Mechanics (2nd edition)} (Addison-Wesley, Reading, MA, 1980).

\bibitem{JS} J.-S. Caux and J. Mossel, J. Stat. Mech. P02023 (2011).

\bibitem{Larson} J. Larson, J. Phys. B {\bf 46}, 224016 (2013).

\bibitem{McGuire} J. B. McGuire, J. Math. Phys. {\bf 5}, 622 (1964).

\bibitem{Yang} C. N. Yang, Phys. Rev. Lett. {\bf 19}, 1312 (1967).

\bibitem{Baxter} R. J. Baxter, Ann. Phys. (N.Y.) {\bf 70}, 193 (1972).

\bibitem{Korepin} V. E. Korepin, N. M. Bogoliubov and A. G. Izergin,  
{Quantum Inverse Scattering and Correlation Functions} (Cambridge University Press, Cambridge, 1997)

\bibitem{Baxterbook} R. J. Baxter, {\em Exactly Solved Models in Statistical Mechanics} (Academic Press, London, 1982).


\bibitem{Braak1} D. Braak,  Phys. Rev. Lett. {\bf 107}, 100401 (2011).

\bibitem{Rabi} I. Rabi,  Phys. Rev. {\bf 51}, 652 (1937). 

\bibitem{qdots} D. Liebfried, R. Blatt, C. Monroe and D. Wineland, Rev. Mod. Phys. {\bf 75}, 281(2003); 
D. Englund {\em et al.}, Nature {\bf 450}, 857 (2007).

\bibitem{sbits} A. Wallraff {\em et al.}, Nature {\bf 431}, 162 (2004); 
I. Chiorescu {\em et al.}, Nature {\bf 431}, 159 (2004).

\bibitem{cavity} J. M. Raimond, M. Brune and S. Haroche, Rev. Mod. Phys. {\bf 73}, 565 (2001). 

\bibitem{circuit} T. Niemczyk {\em et al.},  Nature Phys. {\bf 6}, 772 (2010); 
P. Forn-Diaz {\em et al.},  Phys. Rev. Lett. {\bf 105}, 237001 (2010).

\bibitem{Dicke} R. H. Dicke, Phys. Rev. {\bf 93}, 99 (1954). 

\bibitem{qgate} J. J. Garcia-Ripoll, P. Zoller and J. I. Cirac, Phys. Rev. Lett. {\bf 91}, 157901 (2003); 
Ch. Piltz, B. Scharfenberger, A. Khromova, A. F. Varon and Ch. Wunderlich, Phys. Rev. Lett. {\bf 110}, 200501 (2013).

\bibitem{casanova} J. Casanova, G. Romero, I. Lizuain, J. J. Garc\'ia-Ripoll and E. Solano, 
Phys. Rev. Lett. {\bf 105}, 263603 (2010).

\bibitem{sim} A. Mezzacapo, U. Las Heras, J. S. Pedernales, L. DiCarlo, E. Solano and L. Lamata, 
Sci. Rep. {\bf 4}, 7482 (2014).

\bibitem{largeN} K. Hepp and E. H. Lieb, Ann. Phys. N.Y. {\bf 76}, 360 (1973); 
Y. K. Wang and F. T. Hioe, Phys. Rev. A {\bf 7}, 831 (1973); 
H. J. Carmichael, C. W. Gardiner and D. F. Walls, Phys. Lett. A {\bf 46}, 47 (1973).

\bibitem{EB2003} C. Emary and T. Brandes, Phys. Rev. E {\bf 67}, 066203 (2003).

\bibitem{BGBE} K. Baumann, C. Guerlin, F. Brennecke and T. Esslinger, Nature {\bf 464}, 1301 (2010).

\bibitem{Braak2} D. Braak,  Ann. Phys. Berlin {\bf 525}, L23 (2013); 
J. Phys. A {\bf 46}, 175301(2013). 

\bibitem{Judd} B. R. Judd, J. Phys. C {\bf 12}, 1685 (1979).

\bibitem{Huen1a} H. Zhong, Q. Xie, M. T. Batchelor and  C. Lee, 
J. Phys. A {\bf 46}, 415302 (2013).

\bibitem{Huen1b} A. J. Maciejewski, M. Przybylska and T. Stachowiak, 
Phys. Lett. A {\bf 378}, 16 (2014).

\bibitem{JC} E. T. Jaynes and F. W. Cummings,  Proc. IEEE {\bf 51}, 89 (1963).


\bibitem{review} N. M. Bogoliubov and P. P. Kulish, J. Mathematical Sciences {\bf 192}, 14 (2013).


\bibitem{Peng} J. Peng, Z.-Z. Ren, D. Braak, G.-J. Guo, G.-X. Ju, X. Zhang and X.-Y. Guo, 
J. Phys. A {\bf 47}, 265303 (2014).

\bibitem{Chen2} H. Wang, S. He, L. W. Duan, Y. Zhao and Q.-H. Chen, 
EPL {\bf 106}, 54001 (2014). 

\bibitem{Braak4} D. Braak,  J. Phys. B {\bf 46}, 224007 (2013).

\bibitem{TC} M. Tavis and F. W. Cummings, 
Phys. Rev. {\bf 170}, 379 (1968).

\bibitem{zhou} J. Links, H.-Q. Zhou, R. H.  McKenzie and M. D. Gould, 
J. Phys. A {\bf 36}, R63 (2003).

\bibitem{perk} H. Au-Yang and J. H. H. Perk, in {\em Advanced Studies 
in Pure Mathematics: Proc. Taniguchi Symp. (Kyoto, Oct. 1988)} 
  (Kinokuniya-Academic, Tokyo, 1989) pp 57-94.

\bibitem{Kundu} A. Kundu, 
Phys. Lett. A {\bf 350}, 210 (2006).

\bibitem{EB} C. Emary and T. Brandes, Phys. Rev. A {\bf 69}, 053804 (2004).

\bibitem{ABEB} M. A. Alcalde, M. Bucher, C. Emary and T. Brandes, Phys. Rev. E {\bf 86}, 012101 (2012).

\bibitem{footg} We do not consider the value $g=0$, for which the models trivially decouple.

\bibitem{Frahm} L. Amico, H. Frahm, A. Osterloh, G. A. P. Ribeiro, 
Nucl. Phys. B {\bf 787}, 283 (2007); L. Amico, H. Frahm, A. Osterloh and T. Wirth, 
Nucl. Phys. B {\bf 839}, 604 (2010).


\bibitem{hybrid} P. Treutlein, C. Genes, K. Hammerer, M. Poggio and P. Rabl, in 
{\em Cavity Optomechanics}, M. Aspelmeyer, T. J. Kippenberg and F.  Marquardt (Eds.) (Springer-Verlag, Berlin, 2014) p 327.

\bibitem{Huen2} H. Zhong, Q. Xie, X.-W. Guan, M. T. Batchelor, K. Gao and C. Lee, 
J. Phys. A {\bf 47}, 045301 (2014). 


\bibitem{bt} M. V. Berry and M. Tabor,  Proc. R. Soc. A {\bf 356}, 375 (1977).

\bibitem{Kus} M. Ku\'s, Phys. Rev. Lett. {\bf 54}, 1343 (1985).

\bibitem{EB2003b} C. Emary and T. Brandes, Phys. Rev. Lett. {\bf 90}, 044101 (2003).

\bibitem{AH} A. Atland and F. Haake, New J. Phys. {\bf 14}, 73011 (2012).

 
\bibitem{Chen1} Q.-H. Chen, C. Wang, S. He, T. Liu and K.-L. Wang, 
Phys. Rev. A {\bf 86}, 023822 (2012).

\bibitem{Koc} R. Koc, M. Koca and H. T\"ut\"unc\"uler, J. Phys. A {\bf 35}, 9425 (2002).

\bibitem{BD} C. M. Bender and G. V. Dunne, J. Math. Phys. {\bf 37}, 6 (1996).

\bibitem{QES} A. V. Turbiner, Comm. Math. Phys. {\bf 118}, 467 (1988).

\bibitem{quasi} A. Moroz,  Ann. Phys. (N.Y.) {\bf 338}, 319 (2013). 

\bibitem{zhang} Y.-Z. Zhang, J. Math. Phys. {\bf 54}, 102104 (2013).

\bibitem{footba} Note that Bethe Ansatz like equations have been derived for the Judd points of the Rabi model \cite{zhang}. 
However, these algebraic equations should not be confused with Bethe Ansatz equations typical of YBI models. 


\end{thebibliography}
\end{document}